\documentclass[epj]{svjour}
\usepackage{graphicx,amssymb,amsmath}
\newcommand{\Jpsi}{$J/\psi$ }
\newcommand{\pT}{$p_T$ }

\newcommand{\sNN}{$\sqrt{s_{_{\mathrm{NN}}}}$ }
\newcommand{\s}{$\sqrt{s}$ }
\newcommand{\pp}{$p$+$p$ }
\newcommand{\cucu}{Cu+Cu }
\newcommand{\raa}{$R_{AA}$ }
\begin{document}
\title{Quarkonia Measurements with STAR}
\subtitle{}
\author{Zhangbu Xu \and Thomas Ullrich (for the STAR Collaboration)
\thanks{\emph{website:} http://www.star.bnl.gov}%
}                     
%
%
\institute{Physics Department, Brookhaven National Laboratory, Upton, NY 11973}
\date{Received: date / Revised version: date}
%
\abstract{ We report results on quarkonium production from the STAR
    experiment at the Relativistic Heavy-Ion Collider (RHIC).  \Jpsi
    spectra in \pp and \cucu collisions at \sNN = 200 GeV with
    transverse momenta in the range of 0.5-14 GeV/$c$ and 5-8 GeV/$c$,
    respectively, are presented.  We find that for $p_T > 5$ GeV/$c$
    yields in \pp collisions are consistent with those in minimum-bias
    \cucu collisions scaled with the respective number of binary
    nucleon-nucleon collisions. In this range the nuclear modification
    factor, $R_{AA}$, is measured to be $0.9\pm0.2~(stat)$.  For the
    first time at RHIC, high-\pT $J/\psi$-hadron correlations were
    studied in \pp collisions.  Implications from our measurements on
    \Jpsi production mechanisms, constraints on open bottom yields,
    and \Jpsi dissociation mechanisms at high-\pT are discussed.  In
    addition, we give a brief status of measurements of $\Upsilon$
    production in \pp and Au+Au collisions and present projections of
    future quarkonia measurements based on an upgrades to the STAR
    detector and increased luminosity achieved through stochastic
    cooling of RHIC.   
} 
\maketitle

The dissociation of quarkonia due to color-screening of their
constituent quarks in a Quark-Gluon Plasma (QGP) is a classic
signature for deconfinement in relativistic heavy-ion
collisions~\cite{colorscreen}.  The suppression of the various \Jpsi
and $\Upsilon$ states is determined by their binding energy and the
temperature in the plasma. Results from the PHENIX experiment at RHIC
show that the suppression of \Jpsi as a function of centrality (the
number of participants) is similar to that observed by NA50 and NA60
at the CERN-SPS, even though the temperature and energy density
reached in these collisions is significantly lower than at
RHIC~\cite{colorscreen,RHICIIQuarkonia}. This indicates that at RHIC
energies additional mechanisms countering the suppression, such as
recombination of charm quarks in the later stage of the collision, may
play an important role; they will need to be studied systematically
before conclusions from the observed suppression pattern can be drawn.
In most theoretical models, the dissociation of quarkonia states is
computed at rest relative to the QGP~\cite{colorscreen}. Recently,
techniques based on the AdS/CFT duality have been utilized to study
the dissociation of quark-antiquark pairs with large velocities
relative to the strongly coupled QGP. Calculations in this framework
show that bound states of heavy fermion pairs (an analog of quarkonium
in QCD) have an effective dissociation temperature that decreases with
\pT as $1/\sqrt{\gamma}$~\cite{adscft}, implying an increasing
suppression, \textit{i.e.} decreasing $R_{AA}$, with \pT.  To test
this conjecture measurements of \Jpsi $R_{AA}$ above $p_T>5$ GeV/$c$ are
needed where the effective \Jpsi dissociation temperature is expected
to be lower than the temperatures reached in RHIC collisions
($\sim$1.5 $T_c$)~\cite{adscft,colorscreen,rhicwhitepapers}.

Understanding \Jpsi production requires knowing what what fraction of
$J/\psi$s are produced from \textit{(i)} gluon and heavy-quark
fragmentation, from \textit{(ii)} decay feed-down of B mesons and
$\chi_c$ states, and \textit{(iii)} what fraction originates from
direct production either through a color-octet or color-singlet state.
Our understanding of \Jpsi production mechanisms has gone through
several cycles in recent decades~\cite{CDF,jpsippoverview,belle}.
There is no convincing model explaining all the major features of the
existing data in $e^{+}e^{-}$ annihilation and hadron-hadron
collisions. The color-singlet model (CSM)~\cite{CDF,jpsippoverview}
underpredicts the \Jpsi spectra by an order of magnitude in
$\bar{p}+p$ collisions at \s = 1.8 TeV~\cite{CDF}.  The color octet
(COM) and color evaporation model (CEM) were proposed to explain the
production yields, but fail to explain the recent measurements of
large spin alignment (polarization) from the same
experiment~\cite{CDF,RHICIIQuarkonia}. In addition, \Jpsi is found to
be dominantly associated with open-charm pair production in $e^{+}e^-$
annihilation at \s = 10.6 GeV~\cite{belle}. These later findings
suggest that yields calculated from leading-order pQCD may not be the
dominant contribution to the \Jpsi production. UA1 and D0 have
analyzed the $J/\psi$-hadron correlations in $\bar{p}+p$ collisions to
separate \Jpsi sources at high $p_T$~\cite{UA1} from $\chi_c$ decay
without a near-side correlation and from $B$ hadron decay with a
strong near-side correlation~\cite{UA1}.


In this paper we report on the measurement of \Jpsi production at
midrapidity with the STAR experiment in \pp and \cucu collisions at
\sNN = 200 GeV with transverse momenta in the range of 0.5-14 GeV/$c$
and 5-8 GeV/$c$, respectively.  In addition, $J/\psi$-hadron
correlations, originally proposed and studied by UA1 \cite{UA1} are
presented. The technique used is analog to that deployed in
hadron-hadron correlations studies in STAR \cite{fuqiang}. We will
also briefly report on the status of measurements of the $\Upsilon$ in
\pp and Au+Au collisions and present projections of yields that will
become available after completion of the STAR detector upgrades and
the increase of RHIC luminosity that will be achieved once stochastic
cooling for ion beams is implemented.


\begin{figure}[tbp]
    \centering
    \includegraphics[width=0.40\textwidth]{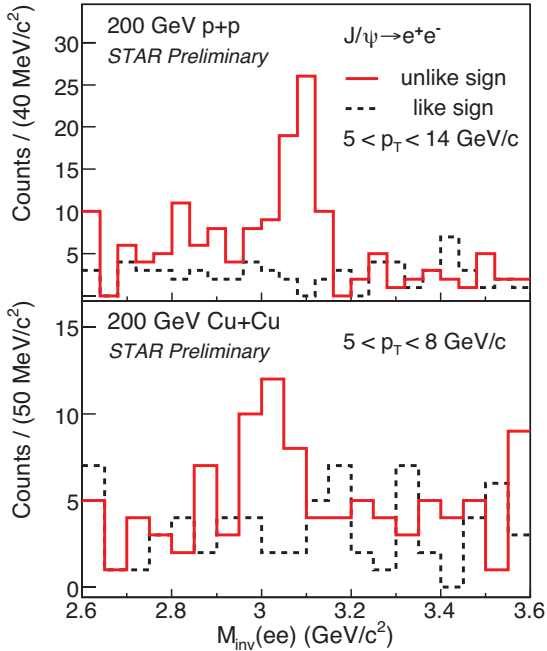}
    \caption{ The $e^+e^-$ invariant mass distribution in \pp
        collisions and Cu+Cu collisions at \sNN =
        200 GeV. The solid and dashed lines represent the
        distribution of un-like and like-sign pairs,
        respectively.}
    \label{invm}
\end{figure}

In the analyses reported here, the \Jpsi and $\Upsilon$ are
reconstructed through their decays into electron pairs, $Q\bar{Q}
\rightarrow e^+e^-$. The large acceptance of the tracking system in
the STAR Time Projection Chamber (TPC)~\cite{stardetector} and the
electron trigger capability from the Barrel Electromagnetic
Calorimeter (BEMC)~\cite{stardetector} with $|\eta|<1$ are very well
suited for such analyses. At STAR, both the TPC and BEMC can provide
electron identification~\cite{stardetector}.  At high $p_T$, the BEMC
is a very powerful tool for electron identification and can be used
for online triggering to enrich the electron sample. At moderate
$p_T$, the TPC provides sufficient energy-loss ($dE/dx$) resolution to
identify electrons efficiently.


Various online trigger schemes were deployed to maximize the recorded
yield.  The low-\pT \Jpsi trigger is based on two spatially separated
BEMC towers with $E_T > 1.2$ GeV.  A higher level trigger makes a
final selection based on the approximated invariant mass of the pair.
However, this trigger can only be efficiently used in \pp collisions;
in A+A collisions this trigger does not provide sufficient
discrimination power due to the high hit occupancy in the BEMC.  The
high-\pT \Jpsi trigger requires only one tower above a given
high-energy threshold. For the data presented here the thresholds
were: $E_T > 3.5$ GeV (\pp run 5), $E_T > 3.75$ GeV (\cucu run 5), and
$E_T > 5.4$ GeV (pp run 6), respectively.  The highly efficient
$\Upsilon$ trigger is based on a single high-tower signal ($E_T > 4.5$
GeV) and a subsequent pair selection with invariant mass cut performed
in an online high-level trigger system using the full BEMC tower data.


For the $\Upsilon$ measurement the sampled luminosities were 9
pb$^{-1}$ in \pp collisions and 12 pb$^{-1}$ (\pp equivalent) in Au+Au
collisions. The low-\pT \Jpsi is from a dataset with 0.4 pb$^{-1}$ of
\pp luminosity.  The data used for the high-\pT \Jpsi analysis was
recorded during the \pp and \cucu runs in 2005 and the \pp run in 2006
using the trigger setup described above in coincidence with a minimum
bias trigger which required a coincidence between the two
Zero Degree Calorimeters (ZDCs). The integrated luminosity was $\sim$
2.8 (11.3) pb$^{-1}$ for \pp collisions collected in year 2005 (2006),
and $\sim$ 860 $\mu$b$^{-1}$ (3 pb$^{-1}$ \pp equivalent) for \cucu
collisions. In \cucu data, the most central 0-60\% of the total
hadronic cross-section was selected by using the uncorrected charged
particle multiplicity at mid-rapidity
($|\eta|<0.5$)~\cite{stardetector,rhicwhitepapers}.


For the high-\pT \Jpsi analysis, we required one high-\pT electron
identified with the combination of BEMC tower energy, shower shape
(from shower-max detectors embedded in the BEMC), and $dE/dx$ measured
in the TPC. The \pT cut for the second electron was substantially
lower and only $dE/dx$ in the TPC was used for identification.

\begin{figure}[tbp]
    \centering
    \includegraphics[width=0.45\textwidth]{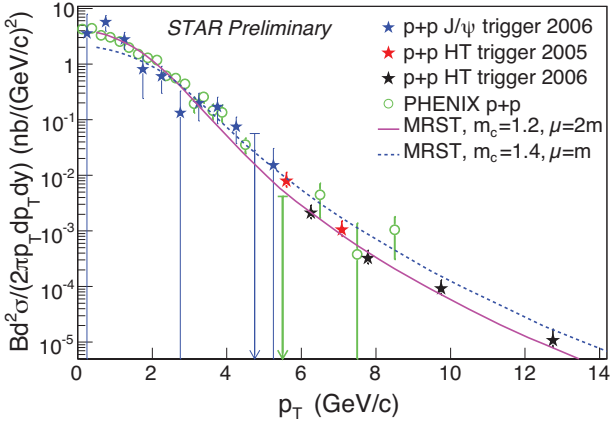}
    \caption{\Jpsi invariant cross-section times the
        di-electron branching ratio as a function of \pT in \pp
        collisions at \s = 200 GeV. The lines correspond to pQCD with
        Color Evaporation Model (CEM) calculations~\cite{RHICIIQuarkonia}.}
        \label{scaling} 
\end{figure}

Figure \ref{invm} shows the invariant mass spectra for the high-\pT
dielectron sample in \pp collisions from runs 5 and 6 (2005 and 2006)
and in Cu+Cu collisions (run 5).  The combinatorial background,
derived from like-sign pairs, is depicted by dashed lines.  The \Jpsi
signal is extracted in the mass window $2.9 < M_\textrm{inv} < 3.2$
GeV/$c^2$. Due to the high-\pT cuts used in this study the
signal-to-background ratio (S/B) is exceptionally large.  In \pp
collisions we obtain S/B = 22/2 (40/14) for run 5 (6) and 17/23 in
\cucu collisions.  The \pT coverage in \pp and \cucu collisions taken
in run 5 is $5<p_T<8$ GeV/$c$.  For \pp collisions in run 6 the larger
integrated luminosity, the larger BEMC coverage, and optimized trigger
thresholds allow us to extend the \pT reach to 14 GeV/$c$.
Figure~\ref{scaling} shows the fully corrected \Jpsi invariant
cross-section $B_{ee}\times Ed^3\sigma/dp^3$ as a function of \pT in
\pp collisions for all three datasets. Our spectrum is in good
agreement with results from the PHENIX experiment, which is plotted
for comparison \cite{phenixpp}.

\begin{figure}[tbp]
    \centering
    \includegraphics[width=0.45\textwidth]{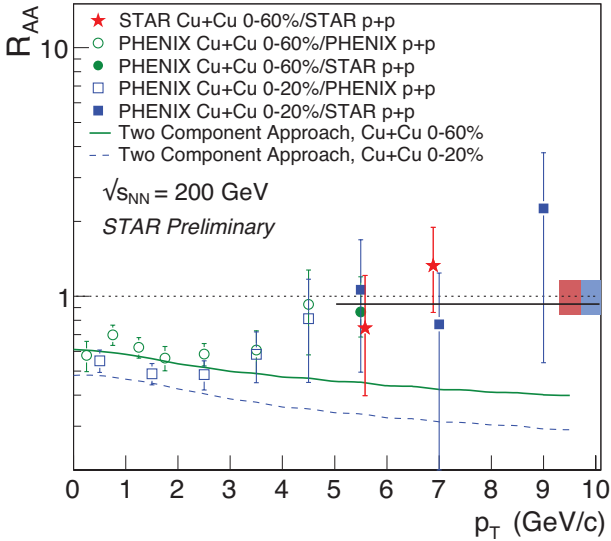}
    \caption{\Jpsi \raa as a function of $p_T$. The solid
         line represents the fit to all the data points at
         5 $<$ \pT$<$ 10 GeV/$c$ yielding $R_{AA}=0.9\pm0.2~(stat)$. 
         The curves are ``Two Component Approach''
         model prediction~\cite{tca}. The boxes in the right show the
         normalization uncertainty for 0-60\% (left) and 0-20\% (right)
         collisions.}
    \label{fig:raa}
\end{figure}

Figure \ref{fig:raa} shows the \Jpsi nuclear modification factors \raa
as a function of \pT in 0-20\% and 0-60\% \cucu collisions calculated
from PHENIX \cite{phenixcucu} and STAR measurements.  The data suggest
that $R_{AA}$ is rising towards unity for $p_T > 5$ GeV/$c$, although
the large errors currently preclude strong conclusions.  A
combined fit to all high \pT data points above 5 GeV/$c$ gives
$R_{AA}=0.9\pm0.2~(stat)$, consistent with unity and $2\sigma$ higher
than that at low-\pT ($R_{AA}\sim0.5$). This result is in
contradiction with expectations from AdS/CFT-based ~\cite{adscft} and
Two-Component-Approach~\cite{tca} models that predict a decrease in
\raa with increasing $p_T$. A similar trend was also observed by NA60
Collaboration in In+In collisions at \sNN $= 17.3$ GeV~\cite{na60}.
There, however, $R_{AA}$ reaches unity at considerably smaller \pT than
at RHIC, suggesting that the effects are most likely of a different
physics origin. Our results could indicate that other \Jpsi production
mechanisms that counter the suppression such as recombination and
formation time effects \cite{csmraa} play an increasingly dominant
role at higher $p_T$. The small suppression of the \Jpsi at high-\pT
stands in contrast to a substantial suppression of open charm
production at similar \pT~\cite{charmsuppression}.


\begin{figure}[tbp]
    \centering
    \includegraphics[width=0.45\textwidth]{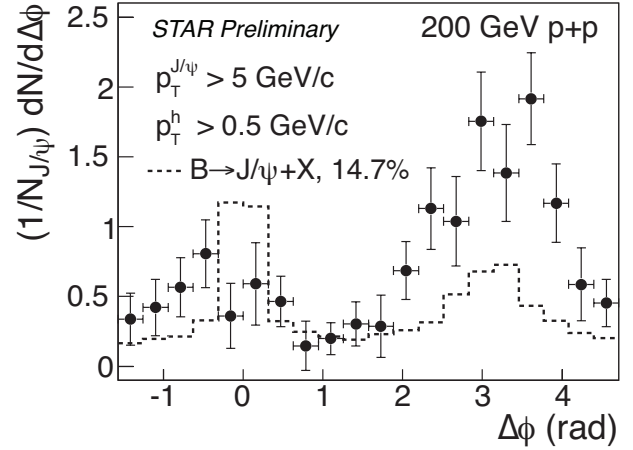}
    \caption{$J/\psi$-hadron azimuthal correlations after
        background subtraction. The histogram is $J/\psi$-hadron from
        B decay with full-event simulation from the PYTHIA event
        generator. }
    \label{corr} 
\end{figure}

The large S/B ratio of the high-\pT \Jpsi data in \pp collisions
allows us, for the first time at RHIC, to study $J/\psi$-hadron
correlations, which can potentially provide important constraints on
the underlying \Jpsi production mechanisms.  Figure~\ref{corr} shows
the azimuthal angle correlations $dN/d\Delta \phi$ per \Jpsi between
$J/\psi$s with $p_T>5$ GeV/$c$ and charged hadrons with $p_T > 0.5$
GeV/$c$. No significant yield for near-side correlations ($\Delta \phi
\sim 0$) is observed. This is in stark contrast to the case of
hadron-hadron correlations~\cite{fuqiang}.

Monte-Carlo simulations (PYTHIA 6.3.19) ~\cite{PYTHIA} show a strong
near-side correlation dominantly due to the feed-down of \Jpsi from
$B$-meson decays \cite{UA1}, $B \rightarrow J/\psi + X$. Thus, the
comparison of the measured near-side yields with PYTHIA simulations
allows one to infer the fraction of $J/\psi$s originating from $B$
meson decays that contribute to the observed yield, albeit in a model
dependent way.  Assuming no other contribution to the near-side, we
obtain an upper limit of $15\%$ of the $B$-meson feed-down to the
inclusive \Jpsi cross-section at $p_T>5$ GeV/$c$. The shape on the
near-side is not matched well.  A detailed comparison of the spectral
shapes is not currently possible due to the limited statistics.  This
will requires data samples which we hope to collect in the upcoming
RHIC run.

It is worth noting that the direct comparison of our inclusive
high-\pT \Jpsi spectra with the feed-down spectra from pQCD
calculations provides less constraints than the method described
above.  Using NLO predictions for $B$ meson production \cite{bcx_pQCD}
and decay parameter from measurements by the CLEO collaboration
\cite{BDecay_CLEO} we obtain an upper limit for the $B$ meson feed-down
fraction of $\sim 40\%$ for $p_T>5$ GeV/$c$. This is in large parts
due to mass and scale uncertainties in the NLO calculations
\cite{RHICIIQuarkonia}.  Further studies of $J/\psi$-hadron
correlation and \Jpsi cross-section with higher statistics will allow
us to constrain the $B$ cross-section substantially in the future.


\begin{figure}[tbp]
    \centering
    \includegraphics[width=0.45\textwidth]{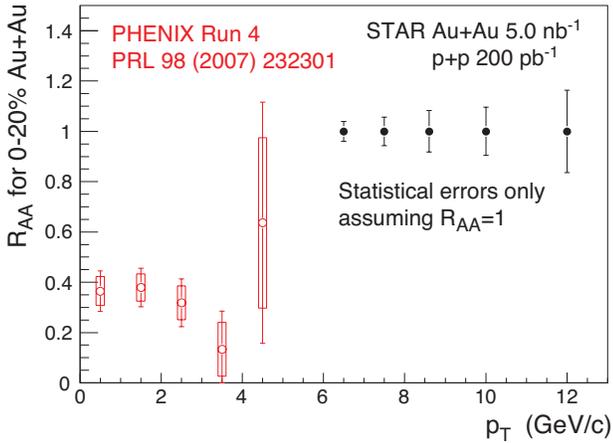}
    \caption{ Projection of $R_{AA}$ at RHIC II
        luminosity, in which STAR will sample 5 nb$^{-1}$ and 200
        pb$^{-1}$ in Au+Au and \pp collisions, respectively. The
        result is compared to $R_{AA}$ at low-\pT obtained by PHENIX
        in run 4, corresponding to a combination of sampled
        luminosities of 157 $\mu$b$^{-1}$ Au+Au collisions and
        2.6 pb$^{-1}$ \pp collisions.}
        \label{jpsifuture}
\end{figure}

Figure~\ref{jpsifuture} shows a projection of $R_{AA}$ for an
integrated luminosity of 5 nb$^{-1}$ for Au+Au collisions and 200
pb$^{-1}$ for \pp collisions with an upgraded STAR detector.  The key
upgrades are: \textit{(i)} the full barrel time-of-flight detector to
improve the electron identification in the low-\pT region ($p_T<3$
$GeV/$c) and \textit{(ii)} the upgrade of the TPC readout electronics
in conjunction with an upgrade of the data acquisition system which
will increase our data-taking rate tenfold and eliminate dead-time
losses for rare triggers. This will allow STAR to make full use of the
increased luminosity of RHIC once the stochastic beam cooling is in
place.

In addition, with the improved electron identification capabilities
and higher data-taking rate the measurement of \Jpsi elliptic flow
becomes feasible. Simulations show that with the statistics of
$10^{9}$ minimum bias Au+Au events we will be able to measure $v_2$ with
$<1\%$ statistical error. This will provide stringent tests of quark
coalescence and charm flow~\cite{RHICIIQuarkonia}.


Many of the difficulties that plague the interpretation of the
observed charmonium suppression can be avoided when studying
bottomonia.  Due to the small production cross-section, recombination
effects become negligible and absorption by hadronic co-moving matter
is unimportant \cite{Lin:2000ke}.  The small cross-section, however,
also makes bottomonium states extremely difficult to measure. While
the $\Upsilon(1S)$ is predicted to be not suppressed at RHIC and even
LHC energies, the $\Upsilon$(2S) ($\Upsilon$(3S)) state is assumed to
dissociate at temperatures similar to that of the \Jpsi
($\psi^\prime$). The separation of the three states, while possible in
STAR, requires substantial statistics which will only become available
with the the RHIC luminosity upgrades.  Detailed $\Upsilon$ simulation
and projections can be found in \cite{RHICIIQuarkonia}.

The STAR experiment already reported on the first RHIC measurement of
the $\Upsilon$(1S+2S+3S) cross-section at mid-rapidity in \pp
collisions at $\sqrt{s} = 200$ GeV. We find $BR \cdot
d\sigma/dy|_{y=0}$ = 91 $\pm$ 28 (stat.) $\pm$ 22 (syst.) pb, which is
consistent with the world data and NLO pQCD calculations in the CEM
\cite{Das:2008nr}.  The first ever measurements of
$\Upsilon$ mesons in Au+Au collisions at $\sqrt{s_{_{NN}}} = 200$ GeV
are underway. We observe a stable signal that will allow us to get
first information on the nuclear modification factor of the $\Upsilon$.
This will be complemented by measurements in d+Au collisions taken in
2008.


In summary, we report measurements of \Jpsi spectra from 200 GeV \pp
up to transverse momenta of 14 GeV/$c$ and from minimum bias \cucu
collisions at high \pT ($5<p_T<14$ GeV/$c$) at mid-rapidity. The \Jpsi
nuclear modification factor \raa in \cucu at \pT$>$ 5 GeV/$c$ is
$0.9\pm0.2~(stat)$ and is about $2\sigma$ above the values at low \pT
measured by PHENIX~\cite{phenixcucu}. The study of \Jpsi-hadron
azimuthal correlations show an absence of near-side correlations.
Using PYTHIA simulations we derive an upper limit for the fraction of
feed-down of \Jpsi mesons from $B$ meson decays of $15\%$ at $p_T>5$
GeV/$c$.  Near future upgrades to the STAR detector and the RHIC
machine will allow us to make substantial contributions to the
understanding of quarkonia production and provide detailed information
of their interaction with the hot dense matter created at RHIC.


\end{document}